\documentclass[a4paper,12pt]{article}
\usepackage{amsmath}
\usepackage{axodraw}

\usepackage{bbm} 
\usepackage[dvips]{hyperref} 
\hypersetup{%
  bookmarksopen=true,
  bookmarksnumbered=true,
  pdftitle = {Z' reconsidered},
  pdfsubject = {Particle physics phenomenology, extra neutral gauge bosons},
  pdfauthor = {A. Ferroglia, A. Lorca, J.J. van der Bij},
}

\usepackage{graphicx}

\newcommand{\urlarxiv}[1]{\href{http://arxiv.org/abs/#1}{[#1]}}
\newcommand{\eq}[1]{Eq.\ (\ref{#1})}

\newcommand{\figs}[2]{Figs.\ (\ref{#1}--\ref{#2})}
\newcommand{\fig}[1]{Fig.\ (\ref{#1})}
\newcommand{\tab}[1]{Tab.\ (\ref{#1})}

\renewcommand{\sec}[1]{section \ref{#1}}

\newcommand{\address}[1]{\begin{minipage}{\textwidth}\begin{center}{\sl #1}\end{center}\end{minipage}}

\def\be#1\ee{\begin{equation}#1\end{equation}}
\def\bea#1\eea{\begin{eqnarray}#1\end{eqnarray}}

\def\im{\mathrm{i}}

\def\afb{A^b_\mathrm{fb}}
\def\ssq{{\sin^2{\theta}}_\mathrm{eff}^\mathrm{lept}}

\def\gev{\textrm{ GeV}}
\def\tev{\textrm{ TeV}}
\def\bml{B\textrm{-}L}
\def\BY{BY}
\def\dd{\mathrm{d}}

\begin{document}
%
%
%
\hspace{\stretch{1}}
{\tt
\begin{tabular}{l}
Freiburg-THEP 06/15
\end{tabular}
}
\vspace{1cm}

\begin{center}
{\boldmath \bf \LARGE 
The $Z'$ reconsidered
\unboldmath
}
\vspace{3em}

{\Large A. Ferroglia\footnote{email: \url{andrea.ferroglia@physik.uni-freiburg.de}}, A. Lorca\footnote{email: \url{alejandro.lorca@physik.uni-freiburg.de}}, and J.J. van der Bij\footnote{email: \url{jochum@physik.uni-freiburg.de}}}
\end{center}
\address{Physikalisches Institut, Fakult\"at f\"ur Mathematik und Physik,\\ Albert-Ludwigs-Universit\"at Freiburg, D-79104 Freiburg, Germany}

\bigskip

%
%



\begin{abstract}
We consider the extension of the standard model with an arbitrary number 
of $U(1)$ gauge fields coupled to baryon-minus-lepton number and/or hypercharge.
Under the assumption that $\afb$ from the LEP1 experiment is an unlucky
 fluctuation, we find moderate evidence for the presence of such fields 
in the precision electroweak data. A relatively large range of the Higgs
boson mass is allowed. We discuss the phenomenology of the extra $U(1)$ fields.
\end{abstract}


\section{Introduction}\label{intro}
The present day data from the high-energy colliders like LEP and the
Tevatron show that almost all data are described by the standard model
at the loop level. Therefore the extensions of the standard model (SM)
tend to be strongly constrained. Typically such extensions will spoil
the agreement with the data through a variety of effects, one of the
most important of which is the appearance of flavor changing neutral
currents. Even the most popular extension, the minimal supersymmetric
extension of the standard model, has to finely tune a number of
parameters. This leaves only one type of extensions that are safe,
namely extensions with singlet particles. Since the discovery that
neutrinos are massive, it is clear that singlet fermions play a role
in nature. There are arguments from cosmology, that singlet
scalars could be important \cite{McDonald:1993ex,Bento:2001yk,Burgess:2000yq,Davoudiasl:2004be,vanderBij:2006ne}. %
In a recent analysis of the Higgs-search
data from LEP2 it has been pointed out, that such singlet scalars may
already have been seen as a smeared-out Higgs boson
\cite{vanderBij:2006pg}. Given this situation it is natural to ask
whether also singlet vector bosons can be present. Therefore we
decided to study the most general renormalizable extension of the
standard model containing extra gauge bosons, but no extra fermions or
scalars. Since the mass of these extra bosons is put in without a
Higgs mechanism, these can only be abelian vector bosons. Demanding
the absence of anomalies in the gauge currents, the extra vector
bosons can only couple to linear combinations of hypercharge ($Y$) and
the difference of baryon and lepton quantum numbers ($\bml$). However they can have an arbitrary mixing with the
standard model hypercharge field. Through this mixing, these fields
introduce small changes in the couplings of the $Z$-boson to
matter. These couplings are suppressed by a factor
$m_Z^2/m_{Z'}^2$. If the masses of  the extra bosons are large enough
such effects are allowed in the data. Of course if the couplings are
infinitesimal, such an extension is always possible. The real question
is whether the effects of the new physics can improve the agreement with the
data. Even though almost all data are described by the SM \cite{lepewwg},
there is the somewhat disturbing situation, that the overall fit to
the standard model is not very good. The reason is the large
difference in $\ssq$ from the forward-backward asymmetry $\afb$ of the
bottom quarks and the measurement from the SLAC SLD experiment. No
realistic model has so far been able to explain this difference. It
has led some authors \cite{Chanowitz:2001bv,Chanowitz:2002cd,Ferroglia:2004jx,Ferroglia:2004tt,Ferroglia:2002rg}, to reanalyze the data leaving
out $\afb$.
In this case one can get a good fit to the data, however with an
unphysical Higgs mass of $50 \gev$. What we will show in this paper, is
that we can get a good fit to the data for a physically allowed Higgs
mass,  if we allow for one or more extra $U(1)$ fields. We must however
emphasize that this works only in the reduced data set without
$\afb$. Using the full data set it appears that one has, close to the
95\% CL, an unresolvable problem. Within the reduced data set we were
able to find confidence level contour plots for the couplings of the
extra vector bosons, as a function of the Higgs mass. One can actually
allow for a larger range of the Higgs boson mass than in the SM.

The outline of the paper is as follows. In \sec{model}, we describe the models. In \sec{fits} we discuss the $Z$-boson couplings in relation to the precision experiments. In \sec{pheno} we discuss the phenomenology of a single $Z'$-boson. In \sec{multiboson} we consider some interesting possibilities with more than one $Z'$-boson. In \sec{conclusion} we recapitulate our conclusions.


\section{The model}\label{model}
There is a large class of models containing extra neutral vector bosons \cite{Leike:1998wr}.
We limit ourselves to the simplest, so-called non-exotic extensions \cite{Appelquist:2002mw,Kors:2005uz,Chang:2006fp,Carena:2004xs}.
In this type of extension the extra gauge bosons couple universally 
to the fermion generations and no exotic fermions are present.
The model consists of the SM plus an extra number 
$n$ of $U(1)$ gauge fields. We call the $U(1)$ fields ${\tilde{C}}_{\mu}^i$,
where $i=0,1,\ldots,n$.
These fields can each have arbitrary couplings to the $Y$ and to $\bml$.
 Other couplings are not possible, because these violate the
renormalizability of the model through the anomaly.
Besides this, we are concerned with the ordinary
$SU(2)_L$ gauge fields $W_{\mu}^a$.
The pure gauge field part of the Lagrangian consists 
of a kinetic and a mass term.

The most general form is:
\begin{equation}
\mathcal{L}_\mathrm{gauge} =  -\frac{1}{4} \sum_{a=1}^3 F_{\mu\nu}^a F_{\mu\nu}^a
-\frac{1}{4} \sum_{i,j=0}^n \tilde{z}_{ij} {\tilde{C}}^i_{\mu\nu} {\tilde{C}}^j_{\mu\nu}
          -\frac{1}{2} \sum_{i,j=0}^n {\tilde{m}}_{ij}^2 {\tilde{C}}_{\mu}^i {C}_{\mu}^j
         ,
\end{equation}
where
 ${\tilde{C}}^i_{\mu\nu}$ and $F^i_{\mu\nu}$  are
the field strength tensors for the $U(1)$s and $SU(2)_L$ fields respectively, and the $\tilde{z}_{ij}$ and $\tilde{m}_{ij}^2$ are the coupling parameters.

Through a linear transformation of fields ($\tilde{C} \rightarrow C$), we can 
bring the Lagrangian in the canonical form in the diagonal mass basis ( \mbox{$\tilde{z}_{ij} \rightarrow \delta_{ij}$},
 \mbox{$\tilde{m}^2_{ij}\rightarrow \delta_{ij} m_i^2$} ).
In order that the photon stays massless, one of the $C_\mu$ fields
must be massless; we will call this field $B_{\mu}\equiv C^0_\mu$.
The Lagrangian becomes therefore:
\begin{equation}
\mathcal{L}_\mathrm{gauge}= -\frac{1}{4} \sum_{a=1}^3 F_{\mu\nu}^a F_{\mu\nu}^a
-\frac{1}{4} B_{\mu\nu}B_{\mu\nu} 
-\frac{1}{4} \sum_{i=1}^n C_{\mu\nu}^i C_{\mu\nu}^i
          -\frac{1}{2} \sum_{i=1}^n m_i^2 C_{\mu}^i C_{\mu}^i.
\end{equation}
Having the fields in the mass basis, the original couplings of the
fields have changed. We have only one extra condition. This is that
the $B_{\mu}$ field must not couple to $\bml$, because otherwise the
photon would have a $\bml$ quantum number ($Q_{\bml}$). The $C_{\mu}^i$ fields have 
couplings $g_Y^i$ to $Y$ and $g_{\bml}^i$ to $\bml$.

The coupling to the fermions is then described through the minimal coupling
\begin{equation}
D_{\mu}\psi = \left( \partial_{\mu} +\im \frac{g}{2} W_{\mu}^a \tau^a_L
   +\im \frac{g}{2 c}\Big( s B_{\mu} + \sum_{i=1}^n g_Y^i C_{\mu}^i \Big) Y 
   +\im \sum_{i=1}^n g^i_{\bml} C_{\mu}^i Q_{\bml} \right) \psi.
\end{equation}
Here $s$ and $c$ are sine and cosine of the weak mixing angle.
The gauge boson coupling to the Higgs field originates from the same covariant derivative, but the Higgs field has $Q_{\bml}=0$. After spontaneous symmetry breaking, we therefore
find a mass matrix of the form:
\begin{equation}
\mathcal{L}_\mathrm{mass}=-\frac{1}{2} \frac{m^2}{c^2} 
     \Big(-c W_{\mu}^3 +s B_{\mu} +\sum_{i=1}^n g_Y^i C_{\mu}^i \Big)^2
-\frac{1}{2} \sum_{i=1}^n m_i^2 C_{\mu}^i C_{\mu}^i.
\end{equation}
We subsequently assume that $m_i \gg m$ and diagonalize the
mass matrix in perturbation theory in the parameters
\begin{equation}
\frac{1}{\mu_i^2}\equiv \frac{m^2}{c^2 m_i^2}.
\end{equation}

In the following we will ignore effects of $\mathcal{O}(1/\mu_i^4)$.

We recover the SM relation between the photon field ($A_\mu$) and the Lagrangian fields 
\begin{equation}
A_\mu = s W^3_\mu + c B_\mu,
\end{equation}
while for the $Z$-boson we find
\begin{equation}
Z_{\mu} = -c W_{\mu}^3 + s B_{\mu} - 
     \sum_{i=1}^n \frac{g_Y^i}{\mu_i^2} C_{\mu}^i,
\end{equation}
with the $Z$-boson mass given by:
\begin{equation}
m_Z^2 = \frac {m^2}{c^2}(1-a_Y),
\end{equation}
where we defined
\begin{equation}
a_Y \equiv \sum_{i=1}^n  \frac{(g_Y^i)^2}{\mu_i^2}.
\end{equation}
For the $Z'$-bosons we find
\begin{equation}
Z^i_{\mu} = C_{\mu}^i +
 \frac{g_Y^i}{\mu_i^2} (-c W_{\mu}^3 + s B_{\mu}),
\end{equation}
with masses
\begin{equation}
m_{Z^i}^2 = m_i^2 \left( 1 + \frac{(g_Y^i)^2}{\mu_i^2} \right).
\end{equation}
With the relations above and $Q=\tau^{(3)}_L/2 +Y/2$
we find for the coupling $\im \bar\psi\gamma_{\mu}g_{\psi}^Z \psi$ 
of the $Z$-boson to the fermions:
\begin{equation}
g_{\psi}^Z = \frac{-g \tau^{(3)}}{4 c}(1-a_Y)
\left(1-4|Q|(s^2- c^2 a_Y) + 4 \tau^{(3)}_L Q_{\bml} a_{\BY} +\gamma_5 \right),
\label{zcoupling}
\end{equation}
where we defined
\begin{equation}
a_{\BY} \equiv \frac{c}{g}
\sum_{i=1}^n \frac{g_Y^i g_{\bml}^i}{\mu_i^2}.
\end{equation}

The isospin quantum number $\tau^{(3)}$ is $+1$ for neutrino and 
up-quark, $-1$ for electron and down-quark.
For the $Z^i$ couplings to the fermions we find
\begin{equation}
g_{\psi}^i = \frac {-g \tau^{(3)}}{4 c} g_Y^i
  \left(1-4|Q|+\gamma_5 + \frac{1}{\mu_i^2}(1-4|Q|s^2 +\gamma_5) \right)
   +g_{\bml}^i Q_{\bml}.
\label{gpsii}
\end{equation}

For the $Z$-pole variables, $a_Y$ and $a_{\BY}$ are sufficient to describe 
the data, but for off-shell quantities also the direct
$Z'$-exchange plays a role. We therefore introduce the quantity
\begin{equation}
a_{\bml} \equiv \frac{c^2}{g^2} \sum_{i=1}^n  \frac{(g^i_{\bml})^2}{\mu_i^2}.
\end{equation}
In the case of a single extra $U(1)$ boson one has the relation
\begin{equation}
a^2_{\BY} = a_Y a_{\bml}.
\end{equation}
When one has more than one extra vector boson this relation is in general not 
true anymore.
\boldmath
\section{$Z$-boson Couplings and Fits to the Precision Data}\label{fits}
\unboldmath

In this section we compare the predictions of the model described above with
the precision measurements of LEP and SLD at the $Z$-boson peak. We restrict our analysis to the case of a single $Z'$-boson. Since $m_{Z'} 
 \gg m_Z$, the $Z$-peak observables are only sensitive to the modified $Z$-boson
couplings. Effects from direct
$Z'$-boson exchange are suppressed by a factor
$\Gamma^2_Z/m^2_{Z'}$.
 The Feynman rule for the vertex coupling the $Z$-boson to a
fermion $f$ is given by
\be
\begin{array}{ccc}
&&\vspace{1em}\\
\begin{picture}(0,0)(0,0)
\SetScale{.8}
%
  \SetWidth{1}
\Photon(-40,0)(0,0){2}{5}
\ArrowLine(40,30)(0,0)
\ArrowLine(0,0)(40,-30)
\end{picture}
 \hspace{10ex}
 &
 \longrightarrow
 &
 \qquad \im \gamma^\mu \left(V_f + A_f \gamma_5 \right),\\
&&\vspace{1em}
 \end{array}
\ee
where the vector and axial couplings $V_f$ and $A_f$ are 
\bea
V_f &=& -\frac{\tau^{(3)}g}{4 c} \left(1 - a_Y - 4 \left|Q\right| \left(s^2 - 
a_Y \right) + 4 a_{\BY} \tau^{(3)} Q_{\bml} \right),
\\
A_f &=& -\frac{\tau^{(3)}g}{4 c}\left(1 - a_Y \right).
\eea
In the equations above, $\tau^{(3)} = \pm 1$ is the isospin third component,
$Q$ the electric charge and $Q_{\bml}$ is $-1$ for leptons and
$1/3$ for quarks.

In perturbation theory, even at the tree-level, in order to make predictions for
physical observables, it is necessary to fix the numerical value of 
the Lagrangian parameters $g$, $s$,
and $m$ in terms of the input parameter set $\alpha$, $m_Z$ and $G_{F}$. In the
model under study, the fitting equations  relating the Lagrangian parameters
with the experimental data of the input parameter set are
\bea
4 \pi \alpha &=& g^2 s^2  , \nonumber \\ 
G_F &=& \frac{1}{4 \sqrt{2}} \frac{g^2}{m^2}  , \\
m_Z^2 &=& \frac{m^2}{c^2} (1-a_Y) .  \nonumber
\eea
By solving the fitting equations, one finds that all the bare Lagrangian parameters
depend on $a_Y$
\bea
g^2 &=& g_*^2 + k_g a_Y, \nonumber \\
m &=& m_* + k_m a_Y,\\
s^2 &=& s_*^2 - k_s a_Y,\nonumber
 \label{solfitting} 
\eea
where the constant introduced above have the numerical values
\be
\begin{array}{r@{~=~}l@{\qquad}r@{~=~}l}
g_*^2 & 0.4322, & k_g & 0.5915,\\
m_* & 80.94 \gev, &  k_m & 55.38 \gev,\\
s_*^2 & 0.2122, &k_s &\frac{c_*^2 s_*^2}{c_*^2 -s_*^2} = 0.2904.
\end{array}
\ee

%
The theoretical predictions for the physical observables have to be expressed in
terms of the quantities $g_*$, $m_*$ and $s_*$ ($c_*^2 = 1-s_*^2$). It is
important to observe that, applying the Eqs.\ (\ref{solfitting}) to the
$Z$-boson coupling in \eq{zcoupling}, one has to operate the replacement
\be
s^2 - c^2 a_Y \quad \rightarrow \quad s_*^2 -  \frac{c_*^4 }{c_*^2 -s_*^2} a_Y.
\ee
In the case in which $Q=-1$, the quantity $1-4 s_*^2$ is close
to zero and therefore sensitive to radiative corrections. This has been well studied
within the SM. The main effect comes from the $\gamma-Z$
mixing terms from fermion loops and leads to a shift in the effective
value of $s_*$ to $s_* \approx 0.2314$, as measured in the fermion coupling to the $Z$-boson, which is the numerical value to use in comparison between the SM and the $Z'$ model.
The remaining uncertainty in the prediction is of the order of the radiative corrections
combined with $Z'$ effects. Given the experimental errors affecting the measurements of the physical observables, the remaining uncertainties on the theoretical side will not affect the comparison with the data.

Numerically we get the following relations
\begin{eqnarray}
R_l\big|_{Z'} &=& R_l\big|_\mathrm{SM} ( 1 + 0.92 a_Y - 1.11 a_{\BY})\\
\sigma_0^\mathrm{had}\big|_{Z'} &=& \sigma_0^\mathrm{had}\big|_\mathrm{SM} (1-0.10 a_Y + 2.28 a_{\BY})\\
\Gamma_Z\big|_{Z'} &=& \Gamma_Z\big|_\mathrm{SM}(1+0.16 a_y - 1.10 a_{\BY})\\
\ssq\big|_{Z'} &=& \ssq\big|_\mathrm{SM} -1.10 a_Y - a_{\BY}\\
m_W\big|_{Z'} &=& m_W\big|_\mathrm{SM} ( 1 + 0.72 a_Y)
\end{eqnarray}

\subsection{Analysis of the Z-pole data}

In order to test whether there is evidence in the data for this
model we used the following precision measurements:
the six $\ssq$ measurements from LEP1 and SLD,
$R_l$, $\sigma^0_\mathrm{had}$ and $\Gamma_Z$ also from LEP1.
Furthermore the  $m_W$ measurements from LEP2 and the Tevatron
were used. $G_F$, $\alpha_\mathrm{em}$ and $m_Z$ are used as the input parameters
of the model. Variation within $1 \sigma$ of $m_t$, $\Delta\alpha_h^{(5)}$ and $\alpha_s$ do not affect significantly the following analysis. To see whether a given model, (i.e. a
prescribed value of $m_H$, $a_Y$ and $a_{\BY}$) fits the data,
we calculated the predicted values of the above mentioned variables
and calculated $\chi^2$ for these values. We varied the parameters
over the physical range $m_H > 115\gev$, $a_Y \geq 0$ and $a_{\BY}$
arbitrary.

We proceed  as follows. We assume that the world
is described by a known value of $m_H$, $a_Y$ and $a_{BY}$
within the physically allowed range. On the basis of these values
a prediction is made for each of the $n$  physical observables $x_i$.
We form the quantity
\begin{equation}
\chi^2 = \sum_{i=1}^n \frac{ \left( x^\mathrm{measured}_i -x^\mathrm{predicted}_i \right)^2}{\sigma_i^2}
\end{equation}
According to statistical theory this quantity should
follow a $P_{n}$ cumulative probability distribution
\begin{equation}
P_n(\chi^2) = \int_{\chi^2} ^\infty \dd y \frac{y^{n/2-1}\,\,e^{-y/2}}{\Gamma(n/2) 2^{n/2}}
\end{equation}

We can calculate the probability $P_n(\chi^2)$.
This number gives the probability, that a random fluctuation in the
data measurements would have a probability
smaller than the actually measured data. If this number is small,
it means that the data are not well described by the assumed values of
$m_H$, $a_Y$ and $a_{BY}$. The 95\% confidence level area in the space of
$m_H$, $a_Y$ and $a_{BY}$ consists of the points with 
$P_n(\chi^2) > 5\% $.
It means that the points outside this range are ruled out with a
``confidence of $95\%$''.

In order to see whether any value of the parameters lies within 95\% CL area, we looked for the lowest value of $\chi^2$. The results are shown in \tab{alldata}. 
For all models the lowest value of $\chi^2$ was found at $m_H=115\gev$,
 giving $\chi^2/\mathrm{d.o.f.} \simeq 19/11$.  This high value 
 indicates a bad fit to the data, as was noticed before in \cite{Chanowitz:2001bv,Chanowitz:2002cd,Ferroglia:2004jx,Ferroglia:2004tt,Ferroglia:2002rg}.
The presence of the extra $U(1)$ field hardly improves the $\chi^2$. This is in
agreement with the result in \cite{Chang:2006fp} for the $a_{\BY}=0$ case. 
Both the SM and the extended models are barely compatible with the data at the 95\% CL.

\begin{table}[!h]
\caption{Test of $\chi^2$ for 11 experimental precision data at 
$m_H=115\gev$.}
\begin{center}
\begin{tabular}{|l||cc|ccc|}
\hline
Model & $\chi^2$ & $P_{11}(\chi^2)$ & $m_H$ & $a_Y$&  $a_{\BY}$ \\
\hline
\hline
SM & 19.1 & 5.9\% & 115 & -- & --  \\
$a_Y$ &19.0 & 6.1\% & 115 & $3.3 \cdot 10^{-5}$ & -- \\
$a_Y,a_{\BY}$ & 19.0 & 6.1\% & 115 & 0 & $4.4 \cdot 10^{-5}$\\
\hline
\end{tabular}
\label{alldata}
\end{center}
\end{table}

There are two possibilities that a bad fit can arise. The first
is that the model is simply not correct. The other possibility 
is that the model is correct, but that the data happen to contain an
unlucky data point that spoils the fit. This last possibility can
 be tested for by removing one measurement from the data and seeing whether there
is a large change in $\chi^2$. If we find a very large change in $\chi^2$
 by removing one point, we consider this as an indication
that this point should not be used in the fit.

We redid the $\chi^2$ analysis removing one of the data points in turn and the results are shown in \tab{alldatabutone} for the more significant changes.

\begin{table}[!ht]
\caption{Test of $\chi^2$ for different sets of 10 data points.}
\begin{center}
\begin{tabular}{|ll||cc|ccc|}
\hline
Excluded & Model & $\chi^2$ & $P_{10}(\chi^2)$ & $m_H$ & $a_Y$&  $a_{\BY}$ \\
\hline
\hline
$A_\mathrm{SLD}$ & SM & 14.8 & 13.9 \% & 134 \gev& --&-- \\
$A_\mathrm{SLD}$ & $a_Y$ & 14.8 & 13.9 \% & 134 \gev & 0 & --\\
$A_\mathrm{SLD}$ & $a_Y,a_{\BY}$ &14.1&16.8 \% & 115 \gev & 0 & $- 1.7  \cdot 10^{-4} $\\
\hline
$\sigma^0_\mathrm{had}$ & SM & 16.2 & \phantom{0}9.5 \% & 115 \gev & -- & -- \\
$\sigma^0_\mathrm{had}$ & $a_Y$ & 16.1 & \phantom{0}9.6 \% & 115 \gev & $3.6 \cdot 10^{-5}$ & --\\
$\sigma^0_\mathrm{had}$ & $a_Y,a_{\BY}$ & 14.0 & 17.1 \% & 194 \gev & $8.9 \cdot 10^{-4}$ & $-7.4 \cdot 10^{-4}$\\
\hline
$\afb$& SM & 13.1 & 21.9 \% & 115 \gev & -- & --  \\
$\afb$& $a_Y$ & \phantom{0}9.8 & 45.7 \% &115 \gev & $3.0 \cdot 10^{-4}$ &-- \\
$\afb$& $a_Y,a_{\BY}$ & \phantom{0}9.5 & 48.8 \% &115 \gev & $1.8 \cdot 10^{-4}$ & $\phantom{+}1.6 \cdot 10^{-4}$\\
\hline
\end{tabular}
\label{alldatabutone}
\end{center}
\end{table}

\begin{figure}[!ht]
\framebox{\includegraphics[scale=1.33]{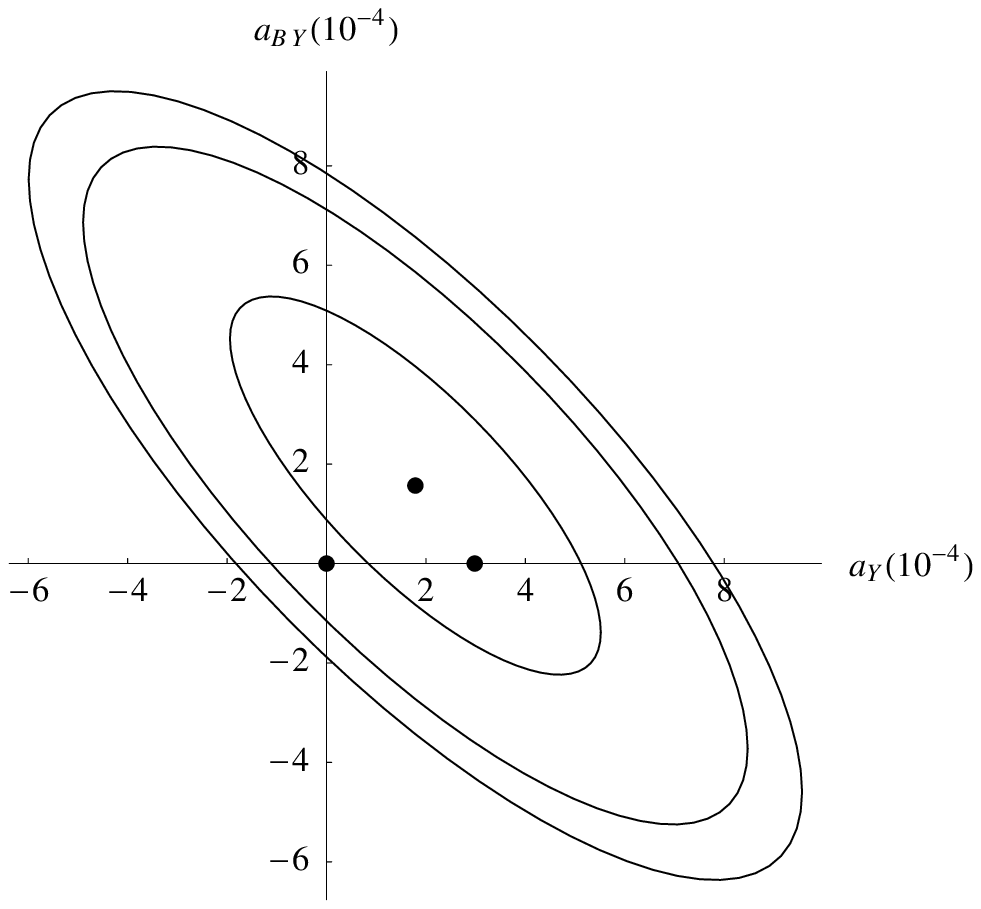}}
\caption{Significance plot for the parameter space \mbox{$ ( a_Y, a_{\BY} ) $} for a fixed \mbox{$m_H = 115 \gev$}.
 The three contours contain a cumulative probability $P_{10}(\chi^2)$ higher than 32\%, 10\% and 5\% corresponding to the 68\%, 90\% and 95\% CL. The point at the origin describes the SM, while the other two points represent the minimum of the $\chi^2$ distribution for the pure $a_Y$ model and for the $a_Y$ plus $a_{\BY}$ model.}
\label{exclusion}
\end{figure}

Only in the case of removing  $\afb$ within the context of the extended models are we able to find a model that cannot be excluded within 68\% CL. Confidence level contours are given in the \figs{exclusion}{higgsmove}.

Altogether we consider this analysis as a mild statistical indication, that the $\afb$ point is an unlucky fluctuation and that the electroweak data are very well
described by the SM with a light Higgs boson and one or more additional
$Z'$-bosons.
We notice that the largest part of the improvement comes from the
presence of the $a_Y$ term, the additional parameter $a_{\BY}$ having a smaller effect.
This allows for models of $Z'$-bosons all coupled to hypercharge only, which is consistent with renormalization.
\vfill
\newpage

\begin{figure}[!ht]
\begin{tabular}{ll}
a) & b)\\
\framebox{\includegraphics[scale=0.63,height=144pt]{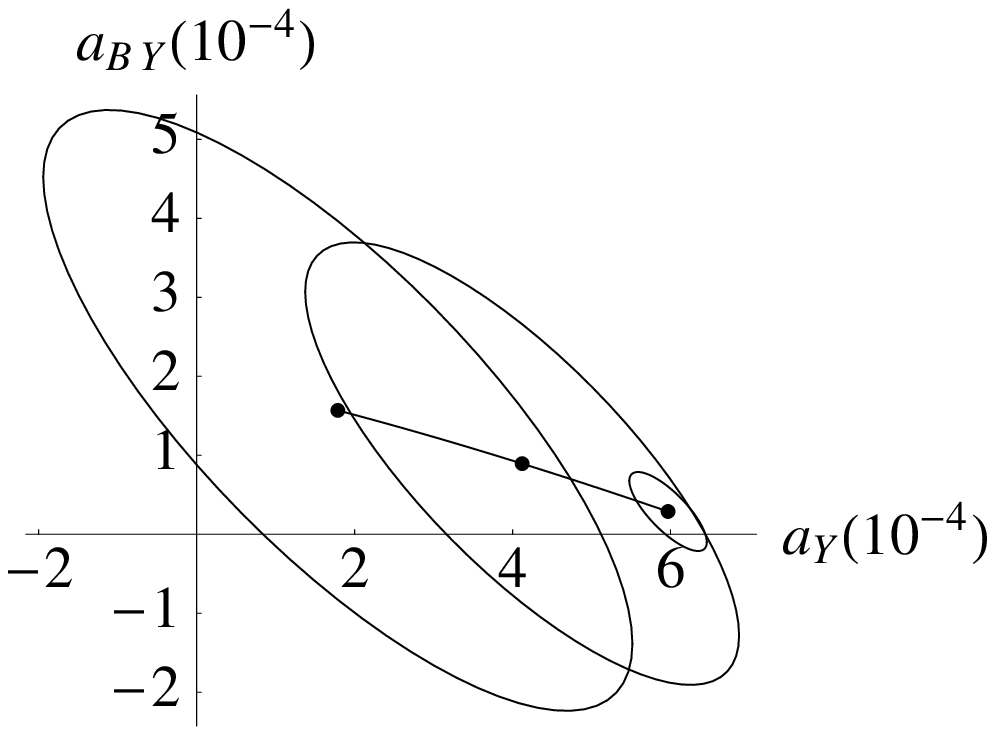}}
&
\framebox{\includegraphics[scale=0.63]{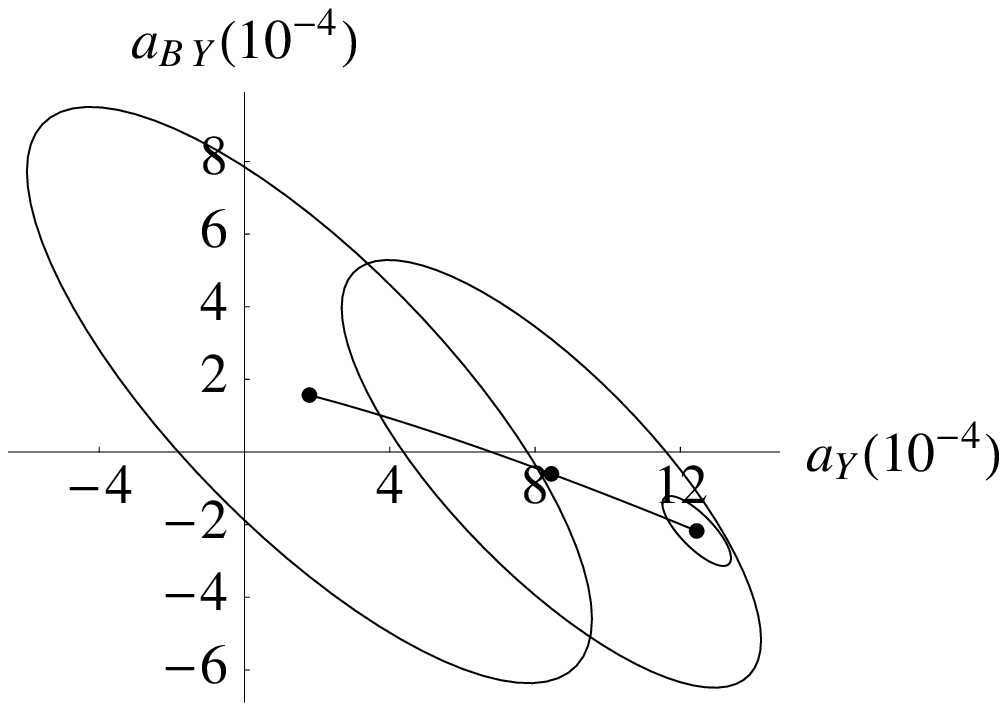}}
\end{tabular}
\caption{Contours of cumulative probability for $P_{10}(\chi^2)$ distribution
 for the parameter space $\{a_Y,a_{\BY}\}$ at different Higgs masses. a) 
The different countours describe 68\% CL and from left to right masses of $115\gev$, $165 \gev$ and $215\gev$. 
The open line moves along the set of minima for the $\chi^2$ up to $m_H=217 \gev$. b) The countours describe 95\% CL and from left to right masses of $115\gev$, $300 \gev$ and $495\gev$. 
The open line moves along the set of minima for the $\chi^2$ up to $m_H=500 \gev$}
\label{higgsmove}
\end{figure}

\subsection{Comparison with off-shell data}
Besides the $Z$-pole observables and the $W$-boson mass, there are a number of
low energy experiments that are in principle sensitive to the
 presence of a $Z'$-boson. Examples are atomic parity violation, the 
NuTeV experiment, M{\o}ller scattering, and  the $(g-2)_{\mu}$ 
experiment. We checked the sensitivity of these experiments
to the $Z'$ model. None of these experiments has a sensitivity to the
$a_Y$ and $a_{\BY}$ variables, that is comparable to the 
sensitivity of the \mbox{$Z$-pole} observables.

Also the LEP2 experiment is sensitive to the presence of
a $Z'$-boson. A recent
analysis of the LEP2 data for lepton production has claimed evidence
for the presence of $Z'$-boson effects \cite{Gulov:2006gv,Gulov:2004sg}. 
Limits were put on the
axial vector and vector couplings of the $Z'$-boson. Translated into
our notation, the following conditions were found:
\begin{eqnarray}
9 a_Y + 24 a_{\BY} + 16a_{\bml} &=& (65 \pm 30 ) \cdot 10^{-3},
\label{lepiiconstraint}\\
a_Y &=& (118 \pm 158 ) \cdot 10^{-3}.
\end{eqnarray}

We see that the errors in these measurements are very large, so one
cannot say that one is testing the model at the same level of precision as
with the \mbox{$Z$-pole} data. The very large value of the couplings 
implied in the first formula
is not very natural compared to the values we derived before. 
In principle,
since we have the new parameter $a_{\bml}$, one could have an extra contribution
coming from a second $Z'$-boson, coupled relatively strongly to ${\bml}$, but
with little mixing. An alternative explanation could be that one is actually
producing directly a spread-out $Z'$-boson. Such a model is presented 
in section 5.
At the moment one should wait for an independent confirmation of the results
in Ref.~\cite{Gulov:2006gv,Gulov:2004sg}.

\boldmath
\section{Phenomenology of $Z'$}\label{pheno}
\unboldmath

\subsection{Renormalization group analysis}\label{rga}
In order to study the phenomenology of the $Z'$-boson
one needs its mass, production cross sections, and 
decay branching ratios. These can be computed when coupling 
constants and mass are known. From the precision data we 
can put limits on the relative values of the $\bml$ and 
$Y$ couplings. But since the limits from the precision
measurements are in the generic form $g^2/m^2_{Z'}$
we have no direct information on the mass itself.
For low masses one can find limits from direct searches.
At first sight there appears to be no way to give an upper limit
to the mass of the $Z'$-boson. A very heavy boson might have a 
large coupling giving similar effects to a light boson with small
couplings. Therefore, to get an upper limit, we need more theoretical input.
Indeed it is the case that one cannot arbitrarily enlarge the couplings.
Since the vector bosons are abelian, their effective couplings
grow via renormalization group and will become infinite 
at the Landau pole. If one demands that the Landau pole is
much larger than the electroweak scale, one can get an upper limit
to the mass of the $Z'$-boson. We will give somewhat qualitative limits
using the one-loop running of the couplings constants due to the fermions 
and demand
that the Landau pole lies beyond the Planck mass scale. In this case
the new $U(1)$ fields would not be too different from the SM gauge fields.
The analysis is slightly complicated, due to the fact that the renormalization
group mixes the different fields \cite{Holdom:1985ag,delAguila:1988jz,delAguila:1995rb}.
We first define the following auxiliary variables 
\begin{equation}
t \equiv \frac{\log(\frac{Q^2}{m_Z^2})}{24\pi^2},  \qquad g_1 \equiv g\frac{s}{c}, \qquad 
g_2 \equiv g\frac{g_Y^1}{c}.
\end{equation}

\begin{figure}[!ht]
\framebox{\includegraphics[scale=1.32]{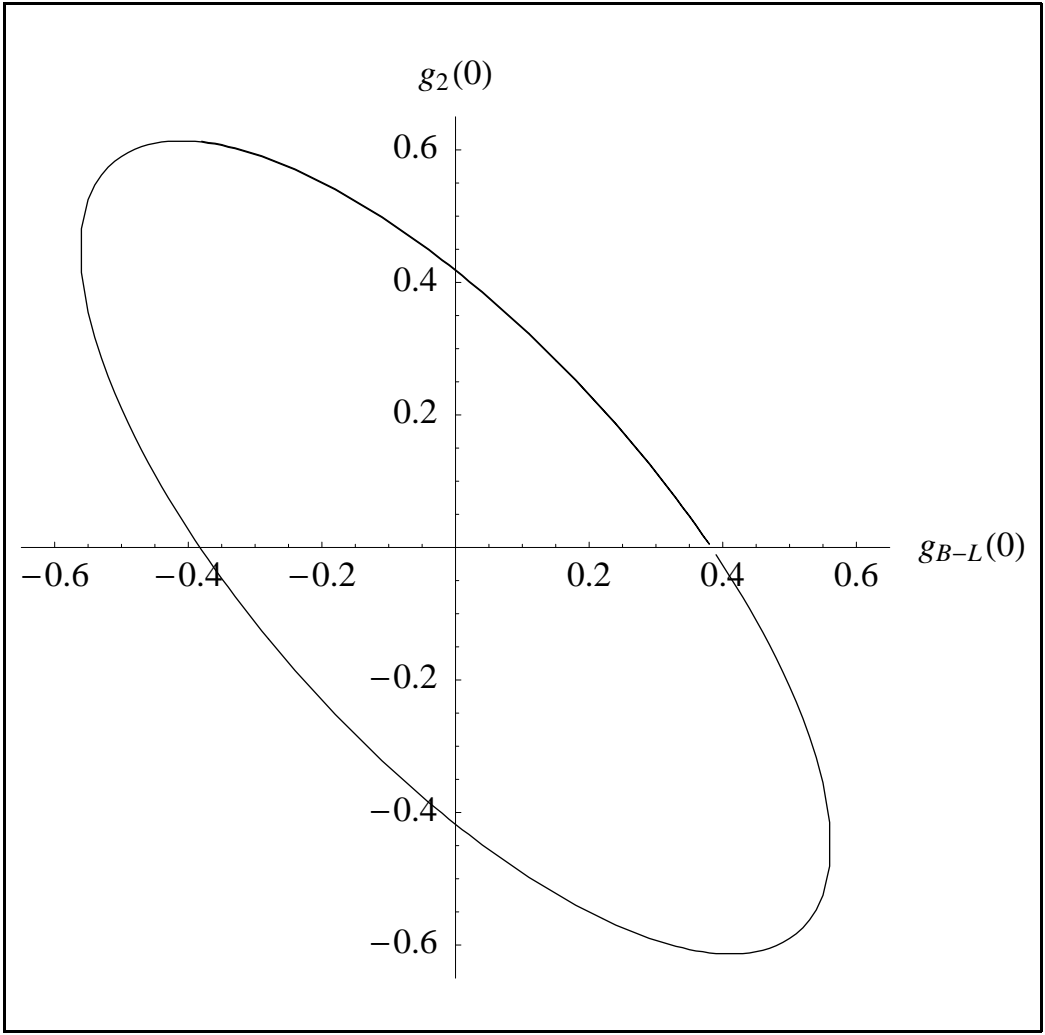}}
\caption{Region of allowed coupling strength at the weak scale.}
\label{renormalization}
\end{figure}

The renormalization group equations become
\bea
\frac{\dd g_1}{\dd t} &=& 5 g_1^3, \label{diffg1}  \nonumber \\
\frac{\dd g_{\bml}}{\dd t} &=& g_{\bml}\left(8 g^2_{\bml} +5 g_2^2 
+ 8 g_2 g_{\bml}\right),\\
\frac{\dd g_2}{\dd t} &=& g_1^2 \left(8 g_{\bml}+10 g_2)
            +g_2 (8 g^2_{\bml}+5 g_2^2+8 g_2 g_{\bml}\right). \nonumber 
\eea
Their solution is given algebraically by using the derived equations
\bea
\frac{5\,g_2 + 4g_{\bml}}{g_{\bml}g_1^2} &=& {\rm constant},\nonumber \\
\frac{5 g_2^2+5 g_1^2-8 g_{\bml}^2}{g_{\bml}^2g_1^2} &=& {\rm constant},\\
\frac{1}{g_1^2} + 10 t &=& {\rm constant}. \nonumber 
\label{solvedg1}
\eea
By demanding that there is no Landau pole before the 
Planck mass we find the allowed region in the
$g_2(0)$ versus $g_{\bml}(0)$ plane given in \fig{renormalization}.

Unfortunately, the constraints we get from this
condition are not very strong. For example, if we take 
the $a_{\BY}=0$ case, one finds the limit \mbox{$g_2(0) < 0.42$}.
This corresponds to a limit 
\begin{equation}
{m_{Z'}^2} < 0.32 \frac{m_Z^2}{a_Y}.
\label{renormlimitay}
\end{equation}
 So, even for the relatively large value of $a_Y = 4 \cdot 10^{-4}$,
one finds a limit of $m_{Z'} < 2.6 \tev $. For smaller values
of $a_Y$ the limit gets correspondingly higher. Therefore, even though
it is plausible that the $Z'$-boson is in the low TeV region
and thereby within reach of the LHC, this is not quite guaranteed.
Stronger upper limits on the $Z'$-boson mass can only be derived,
when more theoretical information is put in, for instance
by demanding some unification of coupling constants.
Assuming \eq{lepiiconstraint} to be true we can get better limits,
due to the effect that in this case the $a_{\bml}$ term is 
the dominant one. Using similar reasoning as above one finds
\begin{equation}
m_{Z'} < 800^{+290}_{-140} \gev.
\end{equation}
In this case the $Z'$-boson could well be within reach of the Tevatron at high luminosity.


\subsection{Collider limits}

Nowadays the best limits on the presence of a $Z'$-boson come from the Tevatron collider. 
The principle guiding the search is straighforward \cite{Abulencia:2006iv}.
One uses the decay of the $Z'$-boson into an electron-positron pair
and looks for a peak in the invariant mass. In addition one can use
information on the forward-backward asymmetry of the leptons. 
One therefore needs a prediction for the production cross section and for the
 branching ratio into leptons. Moreover, when one is working within a specific
model, one can use the distribution of the leptons in the center of 
mass angle $\cos{\theta^*}$, leading to somewhat different limits for different models.
 This search has recently been made at the Tevatron \cite{Abulencia:2006iv} for the class of models discussed in Ref.\ \cite{Carena:2004xs}, but not for the specific models presented here.

In the following we use the narrow width approximation and ignore interference
with $Z$-boson and photon exchange.
The production cross section times branching ratio (BR) is well
described by the formula\footnote{c.f.\ with Eqs.\ (3.8-9) from \cite{Carena:2004xs}}:
\begin{eqnarray}
\sigma(p\bar{p} \rightarrow Z' \rightarrow e^+ e^-) &=& \frac{\pi}{12 s}
 \mathrm{BR}(Z' \rightarrow e^+ e^-) \times \nonumber \\
&&\left( c_u w_u(s,M_{Z'}^2) + c_d w_d(s,M_{Z'}^2) \right),
\end{eqnarray}
where
\begin{equation}
c_u={V'_u}^2+ {A'_u}^2,
\end{equation}
and
\begin{equation}
c_d={V'_d}^2+ {A'_d}^2.
\end{equation}
In the above formulas the $V'$ and $A'$ are the vector and axial-vector couplings 
of the quarks to the $Z'$ as given in \eq{gpsii}.
Furthermore $w_u$ and $w_d$ are the luminosities for up-type quarks and down-type quarks,
$s$ the total energy squared of the collision in the center of mass system.

The predicted branching ratio to electron-positron is given in our model approximately by
\begin{equation}
\mathrm{BR}(Z' \rightarrow e^+ e^-) \simeq \frac{\Gamma(Z' \rightarrow e^+e^-)}{\sum_{f}\Gamma(Z'\rightarrow f\bar{f})}\, ,
\end{equation}
where we have neglected the effect of other than fermionic-pair decays.
For generic massive fermionic-pair decay of a single $Z'$ we have
\begin{eqnarray}
\Gamma(Z' \rightarrow f\bar{f} ) &=& \frac{ m_{Z'}}{12 \pi} \sqrt{1-\frac{4 m_f^2}{m^2_{Z'}}} \times \nonumber \\ 
&&\Bigg[ \Big(1-\frac{m_f^2}{m^2_{Z'}} \Big) ({V'_f}^2 + {A'_f}^2) + 3 \frac{m_f^2}{m^2_{Z'}}({V'_f}^2 - {A'_f}^2) \Bigg].
\end{eqnarray}


At the parton level one can give simple formulas for the forward-backward asymmetry
in the center of mass system.
\begin{equation}
A^{e,q}_\mathrm{fb}= 3\frac{1}{R^{-1}_e +R_e}\frac{1}{R^{-1}_q +R_q},
\end{equation}
where the function $R_f\equiv V'_f/A'_f$ has the following values
\begin{equation}
R_f = \left\{
\begin{array}{r@{,\qquad \textrm{for }f=\;}l}
-3-4\,\frac{a_{\BY}}{a_Y}  &e,\\
-\frac{5}{3}-\frac{4}{3}\,\frac{a_{\BY}}{a_Y} &u,\\
-\frac{1}{3}+\frac{4}{3}\,\frac{a_{\BY}}{a_Y} &d.
\end{array}
\right.
\end{equation}

For $\bar p p$ collisions the measured asymmetry is weighted with the 
quark luminosities. In $p p$ collisions the overall asymmetry disappears
and one must resort to rapidity dependent asymmetries.\\

Now we address the question, whether the limits derived in \sec{fits} can 
be used to make general constraints on the expectations at the Tevatron.
Actually the range of parameters as implied in \fig{higgsmove} is too large
to give much of a prediction even at a 68\% CL. Therefore to get 
an impression of a possible explicit phenomenology we present results
with additional constraints. Actually the analysis of the
precision data has shown, that the biggest improvement comes
from the introduction of the $a_Y$ parameter, the additional
$a_{BY}$ parameter giving a smaller improvement to the fit to the data.
It is therefore reasonable to consider the phenomenology, assuming
$g_{\bml}$ to be absent. This is a consistent condition under renormalization
and forms an interesting class of models by itself.

Since the properties of the $Z'$-boson are strongly correlated
with the mass of the Higgs boson, 
we are interested in knowing what the most likely expectation tells us for 
different Higgs masses. Therefore we also consider the set of models
which move along the $\chi^2$ minimization-line with increasing $m_H$.

In principle the model dependence is largely contained in the parameters
$c_u$ and $c_d$, defined above. As argued in Ref.\ \cite{Carena:2004xs} 
it would be useful to have lower limits on the Higgs-mass presented
in the $c_u - c_d$ plane. Unfortunately such a comparison has not been presented in
the literature. Instead we proceeded estimating the lower limits for the allowed mass of the $Z'$-boson as a function  of either $a_Y$ or $m_H$, depending on the restrictions mentioned above. 
We calculated these ranges as follows: we used a LO program to predict
 the total cross section as a function of the coupling constant and the mass.
 Then, this cross section was normalized to the sequential $Z'$-boson,
 which has a lower limit of $850 \gev$ \cite{Abulencia:2006iv}. 
We then connected the lower bounds of $m_{Z'}$ within the 
 two studied models. The derived bounds are of course somewhat qualitative,
 because the sensitivity to the $\cos{\theta^*}$ distribution 
is not exactly modeled this way. A precise analysis would require 
taking into account a bidimensional distribution, including the
 angular one as well, whereby one cannot ignore detector
effects \cite{Abulencia:2006iv}.
 However this needs a detailed simulation of the detector 
and comparison with the actual data, which is beyond the scope of this paper.
To derive the upper limits we used the results from the previous section.

\begin{figure}[!ht]
\center \framebox{\includegraphics[scale=1.25]{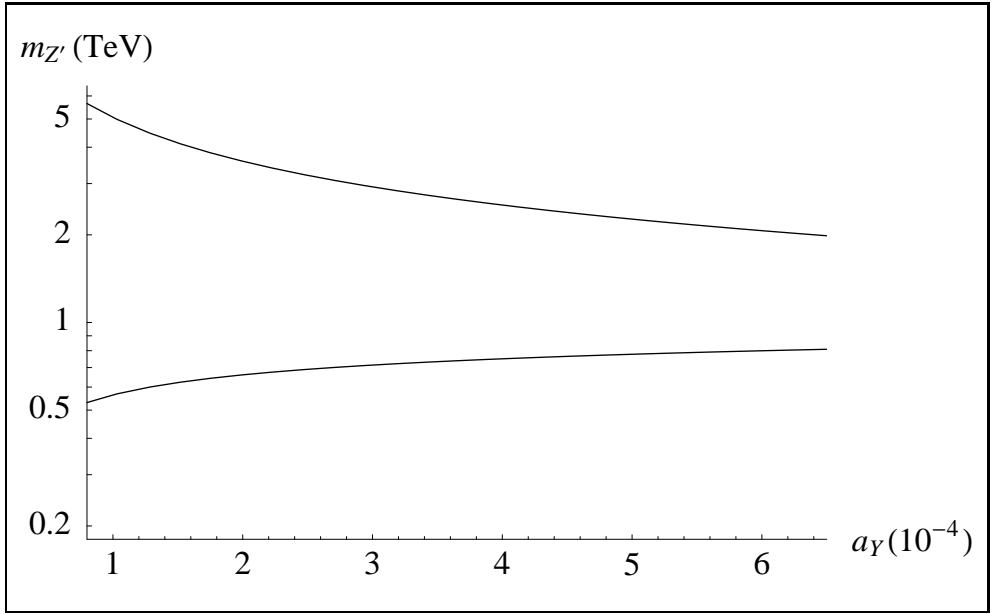}}
\caption{Lower and upper bounds on the $Z'$-mass for models with 
$g_{\bml}=0$.}
\label{zprimemassay}
\end{figure}

In the analysis of the one-parameter model ($a_{\BY}=0$), 
we limit ourselves to the $68\%$ confidence interval
\begin{equation}
0.8\cdot 10^{-4} < a_Y < 6.5 \cdot 10^{-4},
\end{equation} 
which we used to scan the possible lower and upper limits for the $Z'$-boson
 as shown in \fig{zprimemassay}.

As an alternative we considered the best-fitted models for different
values of $m_H$. 
 In these models the parametrized equations used were
\begin{eqnarray}
a_Y(m_H)&=&\left(-114 - 7 \log{(\frac{m_H}{\gev})} + 7.3 \log^2{(\frac{m_H}{\gev})} \right)\cdot 10^{-5},\\
a_{\BY}(m_H)&=& \left(-42.2 + 41 \log{(\frac{m_H}{\gev})} - 6.1 \log^2{(\frac{m_H}{\gev})} \right)\cdot 10^{-5},
\end{eqnarray}
being accurate in the interval
\begin{equation}
115 \gev \le m_H \le 500 \gev.
\end{equation}

Here in order to extract the maxima of the $g_2(0)$ coupling domain,
 we projected the best estimates of $a_Y$ and $a_{\BY}$ onto the
 ellipse of \fig{renormalization}, obtaining thus a non-linear 
mapping $m_H \longmapsto g^1_Y$, which finally can be transformed, 
similarly as in \eq{renormlimitay}, to a $m_{Z'}$ upper limit.

\begin{figure}[!ht]
\center \framebox{\includegraphics[scale=1.25]{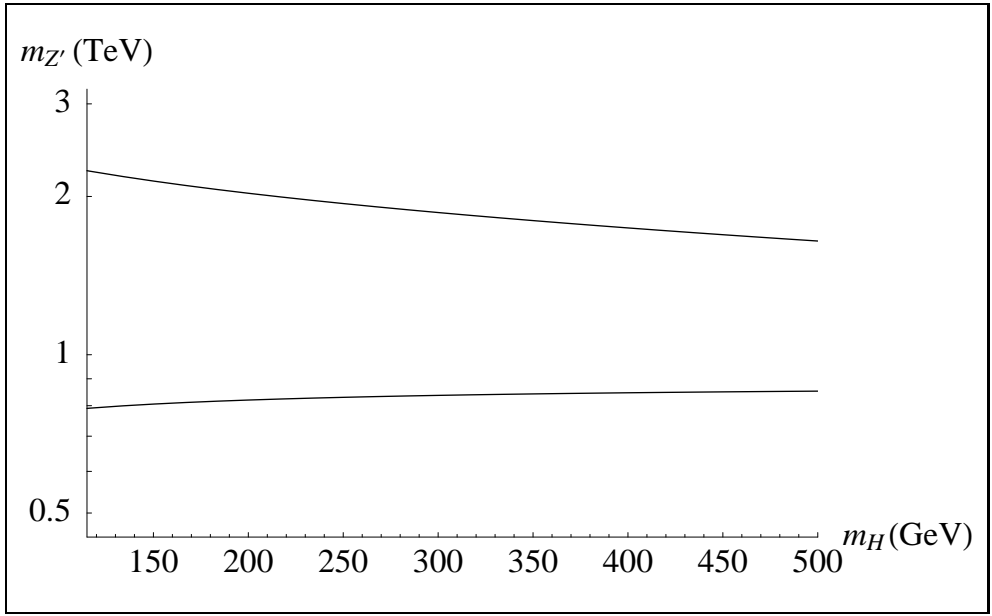}}
\caption{Lower and upper bounds on the $Z'$-mass for the best model values \mbox{$(a_Y,a_{\BY})$} as a function of $m_H$.}
\label{zprimemassmh}
\end{figure}

The result for this analysis is shown on \fig{zprimemassmh}.
In order to avoid a wrong interpretation, we point out that the limits shown in this picture do not correspond to equal values of confidence level at each $m_H$. They just provide the range for the best fit. Actually the confidence limit from the precision data fit ranges between $51\%$ at $m_H=115\gev$ and $95\%$ at $m_H=500\gev$.


Even though the approach is somewhat limited, it suffices to 
retrieve three main features.
First, the Tevatron limits a large part of the a priori allowed masses.
Second, there is a range that could still lead to discovery of a $Z'$-boson 
at the Tevatron with more integrated luminosity. 
Third, we have not enough restrictions on the upper limit to guarantee 
discovery at the LHC. The reason is that $a_Y$ can be quite small. However 
if the Higgs mass is larger than allowed within the standard model fits, 
then $a_Y$ becomes larger and one should be able to find a $Z'$-boson 
at the LHC. For the $Z'$ search at the LHC, the models we discussed
are similar to other models and present no
particular difficulties, therefore we refer to Refs.\ \cite{lhc1,lhc2}.

\boldmath
\section{Multiple $Z'$-bosons}\label{multiboson}
\unboldmath

The formalism as presented in \sec{model} leaves open the possibility
of having more than one extra vector boson.
From the precision data there is no possibility to get information about
the actual number of $Z'$-bosons, because one measures only
the parameters $a_Y$ and $a_{BY}$, that are formed by sums
over all existing vector bosons. If one adds only a finite number
of vector bosons  there will be more $Z'$-bosons with  different sets of
couplings and the renormalization group running becomes more complicated.
However no essentially new aspects appear.
The situation becomes more interesting when one allows for an infinite
number of extra fields. In particular one can also allow for 
vector fields moving in more than four dimensions. However for general
couplings this will lead to a non-renormalizable theory, since 
higher dimensional gauge fields coupled to fermions form operators
of a dimension higher than four. Here the comparison with the data from section
 \ref{fits} gives a useful hint. It was found that the major improvement
in comparison with the data comes from the introduction of the $a_Y$
term. The additional $a_{BY}$ term gives a smaller improvement in $\chi^2$.
It is therefore natural to consider a model where all $U(1)$ fields
couple to hypercharge only. In the mass-basis, that we used so far, we 
then have a large number of fields
all coupled to hypercharge with a certain strength. 
In this case however it maybe advantageous
to study the model in the hypercharge basis. One then has one field coupled
to hypercharge and further fields coupled to nothing. These extra fields make their
presence known only through the mass mixing with the hypercharge field.
Since a mass term for a vector boson has a mass dimension $d-2$ as 
an operator, one can allow for  such vector fields as long as $d \leq 6$.
As an example we therefore take an abelian sector
with a $4$-dimensional field $B_{\mu}$, coupled to hypercharge
and a $d$-dimensional field $A_{\mu}$ coupled to nothing.
We allow for a $4$-dimensional mass term, a mixing mass term 
and a $d$-dimensional mass term.

The Lagrangian becomes
\begin{eqnarray}
\mathcal{L}_\mathrm{gauge} &=& -\int \dd^4x \left( \frac{1}{4} B_{\mu\nu}B_{\mu\nu} + \frac{1}{2} M_4^2 B_{\mu}B_{\mu} \right) \nonumber \\
          && -\int \dd^dx \left( \frac{1}{4} F_{\mu\nu}F_{\mu\nu} +\frac{1}{2} M_5^2 A_{\mu}A_{\mu} \right) \nonumber \\
          && -\int \dd^dx \prod_{i=1}^{d-4}\delta(x_{4+i}) M^{4-\frac{d}{2}}_\mathrm{mix} A_{\mu}B_{\mu}.
\end{eqnarray}
We consider the case that the $A_{\mu}$ fields move in a flat open 
space. By first compactifying the higher dimensions and subsequently
taking the continuum limit one can derive a hypercharge-boson
propagator with a nontrivial K\"all\'en-Lehmann spectral density.
The analysis follows the treatment of the Higgs-singlet
mixing in Ref.\ \cite{vanderBij:2006pg}. For a detailed derivation we refer 
to this paper. Here we only give the result.

The hypercharge propagator becomes of the form
\begin{equation}
D^{BB}_{\mu\nu}(q^2)= \delta_{\mu\nu}
\left[ q^2 +M^2 - \mu_\mathrm{lhd}^{8-d}
(q^2+m^2)^{\frac{d-6}{2}} \right]^{-1}.
\end{equation}

The masses $M$, $m$ and $\mu_\mathrm{lhd}$ are the free parameters of the model.
 The scale $\mu_\mathrm{lhd}$ stands for
low-to-high-dimensional mixing mass and measures the mixing of the high(=$d$) dimensional
vector and the low(=4) dimensional vector. 
The propagator contains a particle peak and a continuum. In order to guarantee
that the particle peak is at mass zero, which is necessary to have a massless
photon in the theory, we must take:

\begin{equation}
M^2 m^{6-d} = \mu_\mathrm{lhd}^{8-d}. 
\end{equation}

We now consider the simple cases of integer dimensions 4, 5 and 6.
In the four dimensional case we should recover the original model.
This is indeed what happens. One finds a single $Z'$-boson, with
\begin{equation}
 m^2_{Z'} = M^2+m^2,
\end{equation}
and 
\begin{equation}
a_Y= \sin^2{\theta_W}\frac{ M^2}{m^2\,(M^2+m^2)}.
\end{equation}

Next we consider the case $d=5$. 
The propagator becomes of the form:

\begin{equation}
D^{BB}_{\mu\nu}(q^2)= \delta_{\mu\nu}\left[ q^2 +M^2 - m M^2
(q^2+m^2)^{-\frac{1}{2}} \right]^{-1} .
\end{equation}

This corresponds to a K\"all\'en-Lehmann spectral density:
\begin{equation}
\rho(s) = \frac{2 m^2}{2 m^2 + M^2} \delta(s) + \frac{\theta(s-m^2)}{\pi} 
\frac{m M^2(s-m^2)^{\frac{1}{2}}}{(s-m^2)(s-M^2)^2+m^2 M^4 }.
\end{equation}

We therefore find one massless particle and a massive continuum.
After spontaneous symmetry breaking the massless excitation
becomes the photon. The massive fields form a spread-out
hypercharge-coupled $Z'$-boson.
As a consequence we find:
\begin{equation}
a_Y = \sin^2{\theta}_W \int_{m^2}^\infty \frac{\dd s}{2 \pi\, m\, s}\,
\frac{(2 m^2+M^2)M^2(s-m^2)^{\frac{1}{2}}}{(s-m^2)(s-M^2)^2+m^2 M^4 }.
\end{equation}

Finally we  consider the case $d=6$.
This case is special, since it corresponds to the limiting dimension, where
the theory is still renormalizable.
Using a limiting procedure 
around $d=6$ the propagator can be written as:

\begin{equation}
D^{BB}_{\mu\nu}(q^2)= \delta_{\mu\nu}\left[ q^2 +M^2 +\mu_\mathrm{lhd}^2
\log{\bigg(\frac{q^2+m^2}{\mu_\mathrm{lhd}^2}\bigg)} \right]^{-1} .
\end{equation}

The spectrum has a massless pole when

\begin{equation}
 M^2 +\mu_\mathrm{lhd}^2 \log{ \left( \frac{m^2}{\mu_\mathrm{lhd}^2}\right)} = 0 .
\end{equation}

The corresponding K\"all\'en-Lehmann spectral density is:
\begin{eqnarray}
\rho(s) = \frac{m^2}{m^2+\mu_\mathrm{lhd}^2}
\delta(s) 
+ \theta(s-m^2)\frac{\mu_\mathrm{lhd}^2}
{\left[s-\mu_\mathrm{lhd}^2\log(\frac{s-m^2}{m^2})\right]^2+\pi^2\mu_\mathrm{lhd}^4} .
\end{eqnarray}
Correspondingly one has:
\begin{equation}
a_Y =\sin^2{\theta}_W \int^{\infty}_{m^2} \dd s \frac{m^2+\mu_\mathrm{lhd}^2}{m^2 \,s}\,
\frac{\mu_\mathrm{lhd}^2}
{\left[s-\mu_\mathrm{lhd}^2\log(\frac{s-m^2}{m^2})\right]^2+\pi^2\mu_\mathrm{lhd}^4}. 
\end{equation}

Since such a spread-out $Z'$-boson has no direct mass peak, its detection
could be quite difficult at hadron colliders. Only an analysis
together with an experimental detector simulation can give a reliable
answer about the limitations in detection here. 
An important role could be played by the correlations for the outgoing leptons.
At the Tevatron such a correlation can be measured directly,
but at the LHC a rapidity dependent analysis will be needed.
It seems even possible, that such a $Z'$ could have been produced
at LEP2, but has been overlooked because of its
rather non-specific signature. A re-analysis of the data at LEP2 with this possibility in mind might be useful.


\section{Conclusion}\label{conclusion}

We studied the class of models of multiple $U(1)$ fields
coupled only to linear combinations of
hypercharge and baryon-minus-lepton number, the so-called
non-exotic $U(1)$ fields. We took a careful look at the
$Z$-pole precision data and argued the case, agreeing
with some considerations in the literature, that one could reasonably
analyze the data without the $\afb$ point. In the 
subsequent analysis we found a moderate indication for the
existence of extra $U(1)$ fields.  We studied the phenomenology
of the extra $Z'$-bosons, concluding that it is possible
that the Tevatron could find the $Z'$-boson and likely,
but not entirely certain, that the LHC will find it.
Furthermore we pointed out the interesting, but somewhat disturbing 
possibility, that the $U(1)$ fields come from higher
dimensions, leading to a spread-out $Z'$-boson, whose study
might be severely limited at a hadron collider.
If the energy would be high enough to produce it, a 
electron-positron collider could study such a spread-out 
signal without severe difficulties.
\bigskip
 
{\bf Acknowledgements} This work was supported by the DFG
through the Graduiertenkolleg ``Physik an Hadron-Beschleunigern''.


\end{document}